\documentclass[aps,prd,preprint,floatfix,nofootinbib,superscriptaddress]{revtex4}
\usepackage{graphicx,subfigure,epsfig,epsf}

\newcommand{\beq}{\begin{equation}}
\newcommand{\eeq}{\end{equation}}
\newcommand{\bea}{\begin{eqnarray}}
\newcommand{\eea}{\end{eqnarray}}

\newcommand{\Lda}{\Lambda}
\newcommand{\g}{\gamma}
\def\Re{{\cal R \mskip-4mu \lower.1ex \hbox{\it e}\,}}
\def\Im{{\cal I \mskip-5mu \lower.1ex \hbox{\it m}\,}}

\def\etal{{\it et al.}}

\def\tev{\,{\ifmmode\mathrm {TeV}\else TeV\fi}}
\def\gev{\,{\ifmmode\mathrm {GeV}\else GeV\fi}}
\def\mev{\,{\ifmmode\mathrm {MeV}\else MeV\fi}}
\def\to{\rightarrow}

\begin{document}

\def\issue(#1,#2,#3){{\bf #1}, #2 (#3)} 

\def\APP(#1,#2,#3){Acta Phys.\ Polon.\ \issue(#1,#2,#3)}
\def\ARNPS(#1,#2,#3){Ann.\ Rev.\ Nucl.\ Part.\ Sci.\ \issue(#1,#2,#3)}
\def\CPC(#1,#2,#3){Comp.\ Phys.\ Comm.\ \issue(#1,#2,#3)}
\def\CIP(#1,#2,#3){Comput.\ Phys.\ \issue(#1,#2,#3)}
\def\EPJC(#1,#2,#3){Eur.\ Phys.\ J.\ C\ \issue(#1,#2,#3)}
\def\EPJD(#1,#2,#3){Eur.\ Phys.\ J. Direct\ C\ \issue(#1,#2,#3)}
\def\IEEETNS(#1,#2,#3){IEEE Trans.\ Nucl.\ Sci.\ \issue(#1,#2,#3)}
\def\IJMP(#1,#2,#3){Int.\ J.\ Mod.\ Phys. \issue(#1,#2,#3)}
\def\JHEP(#1,#2,#3){J.\ High Energy Physics \issue(#1,#2,#3)}
\def\JPG(#1,#2,#3){J.\ Phys.\ G \issue(#1,#2,#3)}
\def\MPL(#1,#2,#3){Mod.\ Phys.\ Lett.\ \issue(#1,#2,#3)}
\def\NP(#1,#2,#3){Nucl.\ Phys.\ \issue(#1,#2,#3)}
\def\NIM(#1,#2,#3){Nucl.\ Instrum.\ Meth.\ \issue(#1,#2,#3)}
\def\PL(#1,#2,#3){Phys.\ Lett.\ \issue(#1,#2,#3)}
\def\PRD(#1,#2,#3){Phys.\ Rev.\ D \issue(#1,#2,#3)}
\def\PRL(#1,#2,#3){Phys.\ Rev.\ Lett.\ \issue(#1,#2,#3)}
\def\PTP(#1,#2,#3){Progs.\ Theo.\ Phys. \ \issue(#1,#2,#3)}
\def\RMP(#1,#2,#3){Rev.\ Mod.\ Phys.\ \issue(#1,#2,#3)}
\def\SJNP(#1,#2,#3){Sov.\ J. Nucl.\ Phys.\ \issue(#1,#2,#3)}


\bibliographystyle{revtex}

\title{$e^+ e^- \to \mu^+ \mu^-$ scattering in the Noncommutative standard model } 




\author{Abhishodh Prakash}
\email[]{abhishodh@gmail.com}
\author{Anupam~Mitra}
\email[]{anupam.mitra@gmail.com}
\author{Prasanta~Kumar~Das}
\email[]{Author(corresponding) : pdas@bits-goa.ac.in, pdasMaparna@gmail.com}

\affiliation{Birla Institute of Technology and Science-Pilani, K. K. Birla Goa campus, NH-17B, Zuarinagar, Goa-403726, India}


\date{\today}

\begin{abstract}
We study muon pair production $ e^+ e^- \to  \mu^+ \mu^-$ in the noncommutative(NC) extension of the standard model using the Seiberg-Witten maps of this to the second order of the noncommutative parameter
$\Theta_{\mu \nu}$. Using $\mathcal{O}(\Theta^2)$ Feynman rules, we  find the $\mathcal{O}(\Theta^4)$ cross section(with all other lower order contributions simply cancelled) for the pair production. The momentum dependent $\mathcal{O}(\Theta^2)$ NC interaction significantly modifies the cross section and angular distributions which are different from the commuting standard model. We study the collider signatures of the space-time noncommutativity at the International Linear Collider(ILC) and find that the process $ e^+ e^- \to  \mu^+ \mu^-$ can probe the NC scale $\Lambda$ in the range $0.8 - 1.0$ TeV for typical ILC energy ranges.
\end{abstract}

\maketitle


\section{Introduction}
The idea that the fundamental scale of gravity can be as low as TeV has drawn a lot of interest among the physics community 
recently. In some brane-world models \cite{ADD98} where this TeV scale gravity is realized, one can principally expect to see some stringy 
effects in the upcoming TeV colliders and in addition the signature of space-time noncommutavity. Interests in the noncommutative(NC) 
field theory arose from the pioneering work by Snyder \cite{Snyder47} and has been revived recently due to developments connected 
to string theories in which the noncommutativity of space-time is an important characteristic of D-brane dynamics at the low energy 
limit\cite{Connes98,Douglas98,SW99}. Although Douglas \etal \cite{Douglas98} in their pioneering work have shown that noncommutative field theory is a well-defined quantum field theory, the question that remains is whether the string theory prediction and the noncommutative effect can be seen at the energy scale attainable in present or near future experiments instead of the $4$-$d$ Planck scale $M_{pl}$. A notable work by Witten \etal \cite{Witten96} suggests that one can see some stringy effects by lowering the threshold value of commutativity to \tev, a scale which is not so far from present or future collider scale. 

~ What is space-time noncommutavity? It means space and time no longer commute with each other. Now writing the space-time coordinates as operators  we find 
\beq 
[\hat{X}_\mu,\hat{X}_\nu]=i\Theta_{\mu\nu}
\label{NCSTh}
\eeq
where the matrix $\Theta_{\mu\nu}$ is real and antisymmetric. The NC parameter $\Theta_{\mu\nu}$ has dimension of area and reflects the extent to which the space-time coordinates are noncommutative i.e. fuzzy. Furthermore, introducing a NC scale  $\Lda$, we rewrite Eq. \ref{NCSTh} as 
\beq 
[\hat{X}_\mu,\hat{X}_\nu]=\frac{i}{\Lda^2} c_{\mu\nu}
\label{NCST}
\eeq
 where $\Theta_{\mu\nu}(=c_{\mu \nu}/\Lda)$ and $c_{\mu\nu}$ has the same properties as $\Theta_{\mu\nu}$. To study an ordinary field theory in such a noncommutative fuzzy space, one replaces all ordinary products among the field variables with Moyal-Weyl(MW) 
\cite{Douglas} $\star$ products defined by
\begin{equation}
(f\star
g)(x)=exp\left(\frac{1}{2}\Theta_{\mu\nu}\partial_{x^\mu}\partial_{y^\nu}\right)f(x)g(y)|_{y=x}.
\label{StarP}
\end{equation}
Using this we can get the noncommutative quantum electrodynamics(NCQED) Lagrangian as
\begin{equation} \label{ncQED}
{\cal L}=\frac{1}{2}i(\bar{\psi}\star \gamma^\mu D_\mu\psi
-(D_\mu\bar{\psi})\star \gamma^\mu \psi)- m\bar{\psi}\star
\psi-\frac{1}{4}F_{\mu\nu}\star F^{\mu\nu} \label{NCL},
\end{equation}
which are invariant under the following transformations 
\bea
\psi(x,\Theta) \to \psi'(x,\Theta) &=& U \star \psi(x,\Theta), \\
A_{\mu}(x,\Theta) \to A_{\mu}'(x,\Theta) &=& U \star A_{\mu}(x,\Theta) \star U^{-1} + \frac{i}{e} U \star \partial_\mu U^{-1},
\eea
where $U = (e^{i \Lambda})_\star$. In the NCQED Lagrangian [Eq.\ref{ncQED}]
$D_\mu\psi=\partial_\mu\psi-ieA_\mu\star\psi$,$~~(D_\mu\bar{\psi})=\partial_\mu\bar{\psi}+ie\bar{\psi}\star
A_\mu$, and $F_{\mu\nu}=\partial_{\mu} A_{\nu}-\partial_{\nu}
A_{\mu}-ie(A_{\mu}\star A_{\nu}-A_{\nu}\star A_{\mu})$. 

 The alternative is the Seiberg-Witten(SW)\cite{SW99,Douglas98,Connes98,Jurco} 
approach in which both the gauge parameter $\Lambda$ and the gauge field $A^\mu$
is expanded as 
\bea \label{swps}
\Lambda_\alpha (x,\Theta) &=& \alpha(x) + \Theta^{\mu\nu} \Lambda^{(1)}_{\mu\nu}(x;\alpha) + \Theta^{\mu\nu} \Theta^{\eta\sigma} \Lambda^{(2)}_{\mu\nu\eta\sigma}(x;\alpha) + \cdot \cdot \cdot \\
A_\rho (x,\Theta) &=& A_\rho(x) + \Theta^{\mu\nu} A^{(1)}_{\mu\nu\rho}(x) + \Theta^{\mu\nu} \Theta^{\eta\sigma} A^{(2)}_{\mu\nu\eta\sigma\rho}(x) + \cdot \cdot \cdot
\eea
and when the field theory is expanded in terms of this power series Eq. (\ref{swps}) one ends up with an infinite tower of higher dimensional operators which renders the theory nonrenormalizable. However, the advantage is that this construction can be applied to any gauge theory with arbitrary matter representation. In the WM 
approach the group closure property is only found to hold for the $U(N)$ gauge theories and the matter content is found to be in the (anti)-fundamental and adjoint representations. 
Using the SW-map, Calmet \etal \cite{Calmet} first constructed a model with noncommutative gauge invariance which was close to the usual commuting  standard model(CSM) and is known as the {\it minimal} noncommutative standard model(mNCSM) in which they listed several Feynman rules comprising NC interaction. Intense phenomenological searches \cite{Hewett01} have been made to unravel several interesting features of this mNCSM. Hewett \etal explored several processes e.g. 
$e^+ e^- \to e^+ e^-$ (Bhabha), $e^- e^- \to e^- e^-$ (M\"{o}ller), 
$e^- \g \to e^- \g$, $e^+ e^- \to \g \g$ (pair annihilation), $\g \g \to e^+ e^-$ and $\g \g \to \g \g$ in the context of NCQED and NCSM. 
Recently, one of us has investigated the impact of $Z$ and photon exchange in the Bhabha and the M\"{o}ller scattering, which is reported in \cite{pdas}.  Conroy \etal \cite{Conroy} have investigated the process $e^+ e^- \to \gamma \to \mu^+ \mu^-$ in the context of NCQED and predicted a reach of $\Lda = 1.7$ TeV. In addition to the photon($\gamma$) exchange, we also consider the $s$-channel exchange of the $Z$ boson.  Now in a generic NCQED the triple photon vertex arises to order 
${\mathcal{O}}(\Theta)$, which however is absent in this mNCSM. Another formulation of the NCSM came to the  forefront through the pioneering work by  Melic \etal \cite{Melic:2005ep}
where such a triple neutral gauge boson coupling \cite{Trampetic} appears naturally in the gauge sector. We will call this the nonminimal version of NCSM or simply NCSM. The Feynman rules to order $\mathcal{O}(\Theta)$ were presented in their work  \cite{Melic:2005ep}. In 2007, Alboteanu {\it et al} presented the $\mathcal{O}(\Theta^2)$ Feynman rules for the first time. In the present work we will confine ourselves within this nonminimal version of the NCSM and use the Feynman rules given in Alboteanu  \etal \cite{Ana}.

 In Sec. II we present the cross section of $e^+ e^- \to \gamma, Z \to \mu^+ \mu^-$. The numerical analysis and the prospects of TeV scale noncommutative geometry are discussed in Sec. III. Finally, we summarize our results in Sec. IV.  
\section{$ e^+ e^-\rightarrow \mu^+ \mu^-$ scattering in the NCSM}
\begin{figure}[htbp]
\vspace{5pt}
\centerline{\hspace{-3.3mm}
{\epsfxsize=14cm\epsfbox{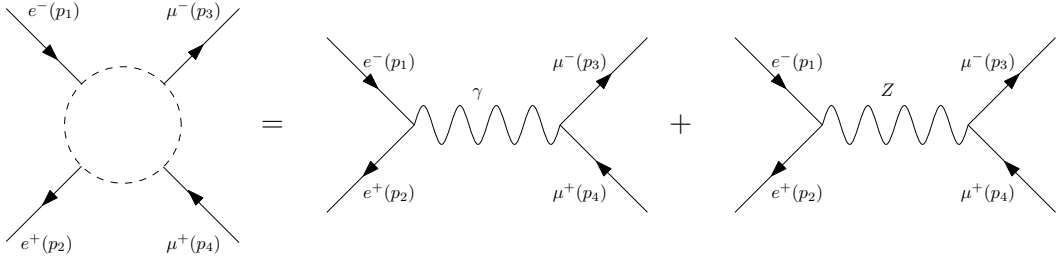}}}
\hspace{2.5cm}
\vspace*{-0.5in}
\caption{Feynman diagrams for $ e^+ e^-\rightarrow \gamma,Z \rightarrow \mu^+ \mu^-$ in the NCSM.}
\protect\label{feyn}
\end{figure}

The muon pair production process $e^- (p_1) e^+ (p_2) \rightarrow  \mu^- (p_3) \mu^+ (p_4)$ proceeds via the $s$ channel exchange of $\gamma$ and $Z$ bosons in the NCSM, like the standard model. The corresponding Feynman diagrams are shown in Fig. \ref{feyn}. 

\noindent In order to have the cross section to order $\mathcal{O}(\Theta^2)$, we include the order $\mathcal{O}(\Theta^2)$ Feynman rule. The scattering amplitude to order $\Theta^2$ for the photon mediated diagram can 
be written as  
\bea \label{gamma}
{\mathcal{A}}_\gamma = \frac{4 \pi \alpha}{s} \left[{\overline v}(p_2) \gamma_\mu u(p_1)\right]  
\left[{\overline u}(p_3) \gamma^\mu v(p_4)\right]  \times
 \left[(1 - \frac{(p_2 \Theta p_1)^2}{8}) + \frac{i}{2} (p_2 \Theta p_1)\right] \nonumber \\ \times \left[(1 - \frac{(p_4 \Theta p_3)^2}{8} ) + \frac{i}{2} (p_4 \Theta p_3)\right]
\eea
and the same for the $Z$ boson mediated diagram as 
\bea \label{Z}
{\mathcal{A}}_Z = \frac{\pi \alpha}{\sin^2(2\theta_W) s_Z}  \left[{\overline v}(p_2) \gamma_\mu (a + \gamma^5)  {\overline u}(p_1)\right] \times \left[{\overline u}(p_3) \gamma^\mu (a + \gamma^5)  {\overline v}(p_4)\right] \nonumber \\ \times  \left[(1 - \frac{(p_2 \Theta p_1)^2}{8}) + \frac{i}{2} (p_2 \Theta p_1)\right] \times \left[(1 - \frac{(p_4 \Theta p_3)^2}{8} ) + \frac{i}{2} (p_4 \Theta p_3)\right]
\eea

\noindent where $s=(p_1 + p_2)^2$, $\alpha = e^2/4\pi$ and $\theta_W$ is the Weinberg angle, $a = 4 \sin^2(\theta_W) - 1$. In the above
$s_Z = s - m_Z^2 - i m_Z \Gamma_Z$,  where $m_Z$ and $\Gamma_Z$ are the mass and decay width of the $Z$ boson. The Feynman rules required for the above scattering process are listed in Appendix A.

\noindent The spin-averaged squared-amplitude is given by
\beq \label{Ampsq}
{\overline {|{\mathcal{A}}|^2}} = {\overline {|{\mathcal{A}}_\gamma|^2}} + {\overline {|{\mathcal{A}}_Z|^2}} + 2 {\overline {Re({\mathcal{A}}_\gamma {\mathcal{A}}_Z^{ \dagger })}}. 
\eeq
The different terms in the spin- averaged squared-amplitude are given in Appendix C. We use the Feynman rule to order ${\mathcal{O}}(\Theta^2)$ while calculating several squared-amplitude and interestingly we found that all lower order terms, i.e. ${\mathcal{O}}(\Theta)$, ${\mathcal{O}}(\Theta^2)$, and ${\mathcal{O}}(\Theta^3)$, get canceled (see Appendixes C and D for further discussions). With these the differential cross section can be written as
\beq \label{dsigma}
\frac{d \sigma}{d \Omega} = \frac{1}{64 \pi^2 s} {\overline {|{\mathcal{A}}|^2}}, 
\eeq
where $\sigma$ = $\sigma(\sqrt{s}, \Lambda, \theta, \phi)$. From Eq. \ref{dsigma} we can obtain $\sigma$, $ d\sigma/d\cos\theta $ and $ d\sigma/d\phi $ as
\bea 
\label{sigma}
\sigma &=& \int_{-1}^1 d(\cos\theta) \int_0^{2 \pi} d\phi \frac{d \sigma}{d \Omega}, \\
\label{dsdcostheta}
\frac{d\sigma}{d\cos\theta} &=& \int^{2 \pi}_0 d\phi \frac{d \sigma}{d \Omega},  \\
\label{dsdphi}
\frac{d\sigma}{d\phi} &=& \int^1_{-1} d(\cos\theta) \frac{d \sigma}{d \Omega}. 
\eea

\section{Numerical Analysis}
In this section, we analyze the total cross section and angular distributions of the differential cross section in the presence of space-time non commutativity obtained in the earlier section. Before making a detailed analysis, let us make some general remarks regarding the observation of noncommutative effects. Since we assume $c_{\mu\nu}=(\xi_i,~\epsilon_{ijk}\chi^k)$, 
where $\xi_i = (\vec{E})_i$ and $\chi_k = (\vec{B})_k$ are constant vectors in a frame that is stationary with respect to fixed stars, the vectors $(\vec{E})_i$ and $(\vec{B})_k$ point in fixed directions which are the same in all frames of reference. However, as the Earth rotates around its axis and revolves around the Sun, the direction of $\vec{E}$ and $\vec{B}$ will change continuously with time dependence which is a function of the coordinates of the laboratory. The observables that are measured will thus show a characteristic time dependence. It is important to be able to measure this time dependence to verify such noncommutative  theories. In our analysis, we have assumed the vectors $\vec{E}= \frac{1}{\sqrt{3}} (\hat{i} + \hat{j} + \hat{k}) $ and $\vec{B}= \frac{1}{\sqrt{3}} (\hat{i} + \hat{j} + \hat{k})$ i.e. they behave like constant vectors. This can be true only at some instant time at most.

\subsection{Production cross section vs the machine energy in the NCSM}
In Fig.2  ~we show the total cross section $\sigma(e^- e^+ \to \mu^- \mu^+)$ as a function of the center-of-mass energy $E_{com}(=\sqrt{s})$ (GeV). The lowermost solid curve(in each of the two figures) corresponds to the CSM (recovered from the NCSM in the $\Lambda \longrightarrow \infty $ limit), whereas the uppermost (long-dashed) curve, next to the uppermost (short-dashed) and next-to-next uppermost (i.e. dotted) curves, arises in the NCSM with $\Lambda = 800,~900,$ and $1000$~GeV, respectively. 
We observe that although the deviation from the commuting standard model is small at relatively lower energies, it starts becoming  significant at $\sim 1400$ GeV and becomes more pronounced with the increase in machine energy. 
Moreover we can also see that at a given machine energy $E_{com}$, the deviations become larger with smaller values of $\Lambda$. 

\begin{figure}[htbp]
\vspace{-1.15in}
\centerline{\hspace{-12.3mm}
{\epsfxsize=9cm\epsfbox{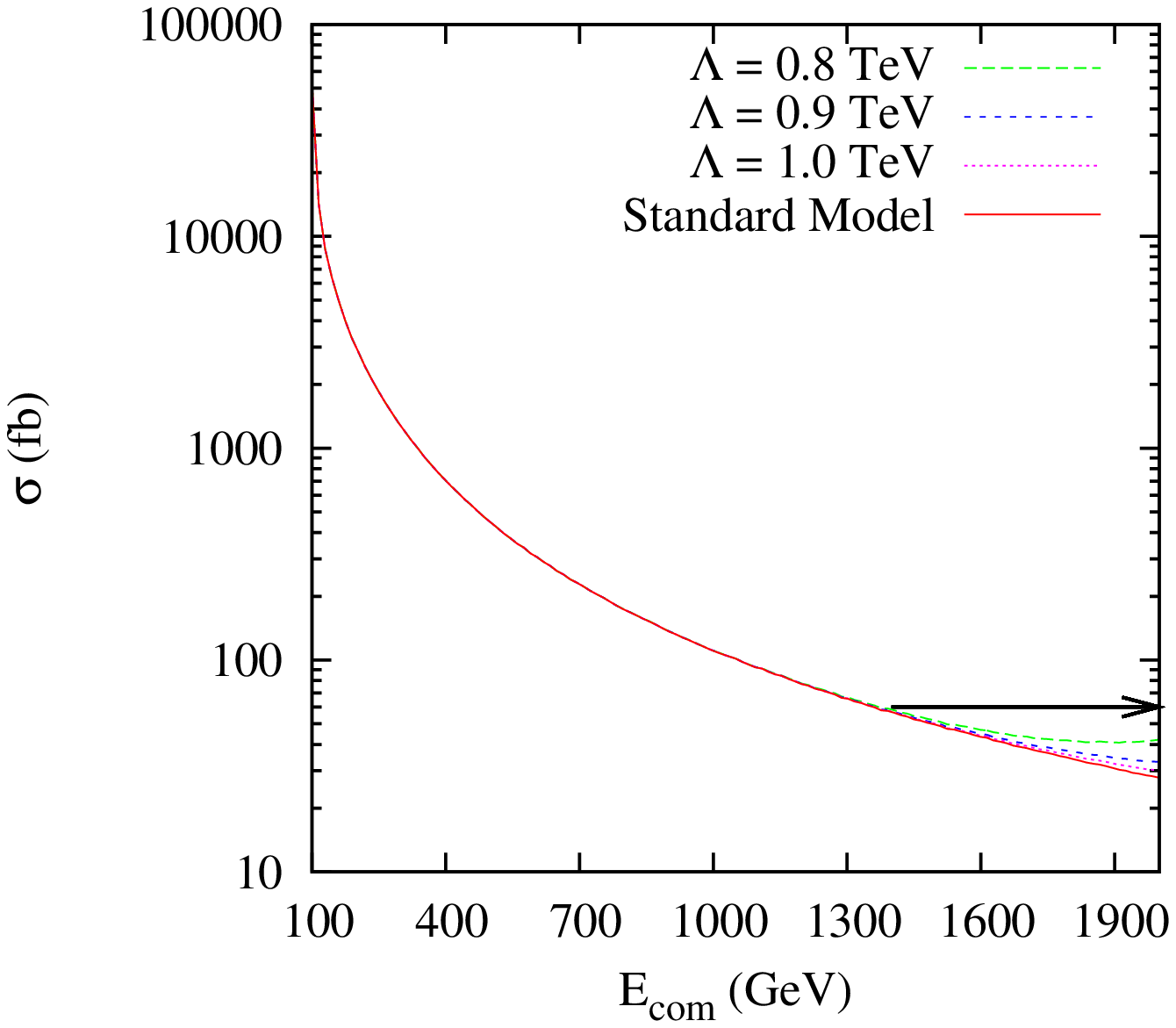}} \hspace{-0.25in} {\epsfxsize=9cm\epsfbox{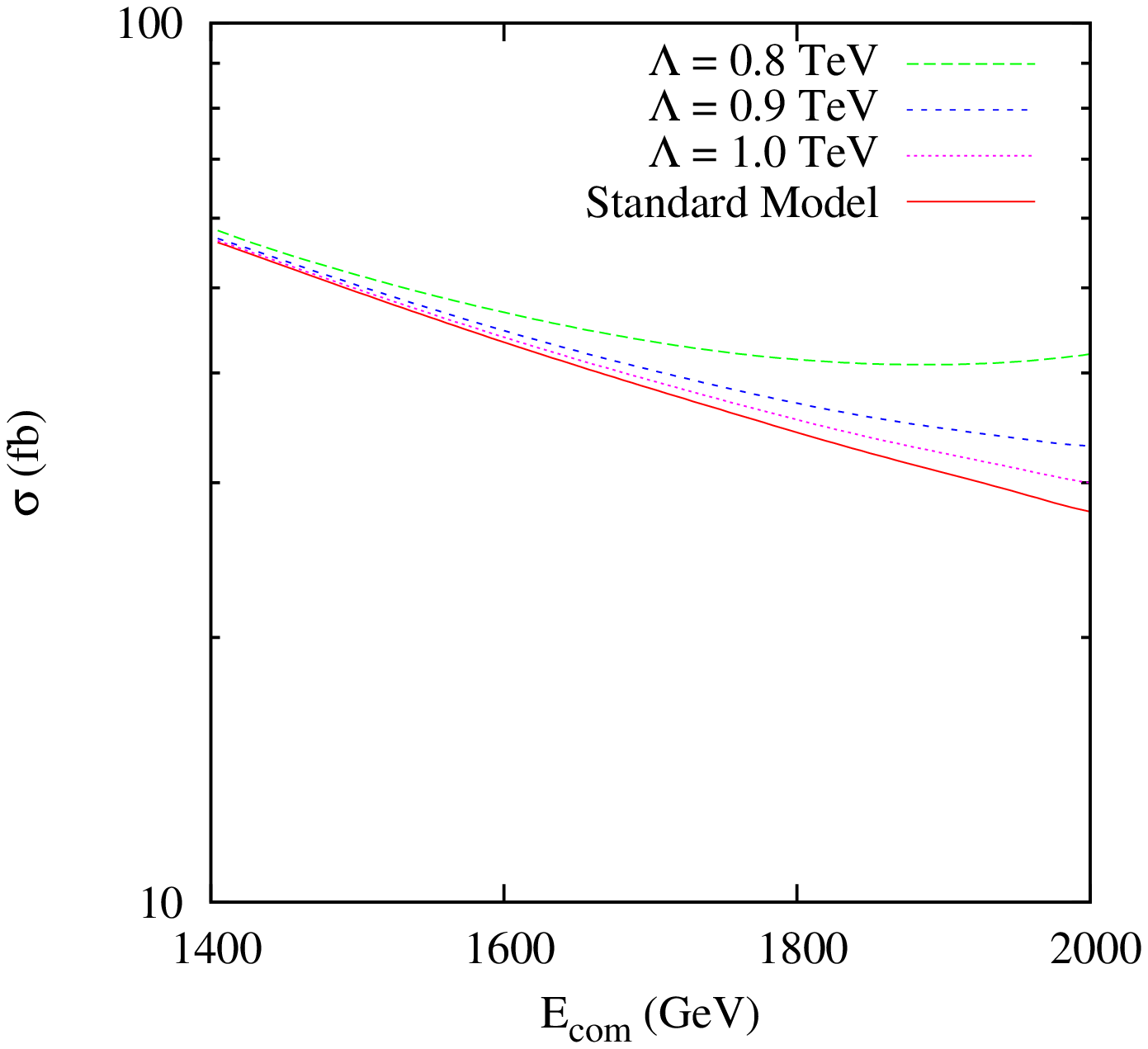}} }
\vspace*{-1.25in}
\caption{The cross section $\sigma(e^- e^+ \to \mu^- \mu^+)$ (fb) as a function of the machine energy $E_{com}=\sqrt{s}$ (in GeV). The figure on the right corresponds to $\sqrt{s} \ge 1400$~GeV. }
\protect\label{sigplot}
\end{figure}
\vspace*{-0.25in}
We made an estimate of the number of events per year($yr^{-1}$) in the case of ILC (International Linear Collider).  
Assuming that the ILC will run for a year with the integrated luminosity $\mathcal{L} = 100~fb^{-1}$, the number of events $N_{SM}~$($yr^{-1}$) at $\sqrt{s} = 1750$ GeV in the CSM is expected to be 
$N_{SM}(=\sigma \times {\mathcal{L}} = 36 \times 100) = 3600$ $yr^{-1}$.  The expected number of events(signals)in the NCSM are given in Table 1. Fixing the machine energy $E_{cm}$ at $1750$ GeV, if we lower $\Lambda$ from $900$ GeV to $800$ GeV, the number of NC events $N$ per year increases from $3800$ $yr^{-1}$ to  $4200$ $yr^{-1}$, which is larger than $N_{SM}(=3600~yr^{-1})$. Note that the NC signal is always larger than the SM background.
\newpage
\begin{center}
Table 1
\end{center}
\begin{center}
\begin{tabular}{|c|c|c|c|}
\hline
$\Lambda$ & NC signal ($\sigma$)(fb) & ${\mathcal{L}}(fb^{-1})$  & N (events per year)\\
\hline
\hline
  800  & 42 & 100 & 4200 \\
\hline
  900 & 38 & 100 & 3800 \\
\hline
  1000 & 37 & 100 & 3700 \\
\hline
\end{tabular}
\end{center}
\noindent {\it Table 1: Progressive reduction of the NC signal and the number of events per year with the increase in the NC scale $\Lambda$.  The machine energy is fixed at $E_{com} = 1750$ GeV. The integrated luminosity of the ILC is assumed to be ${\mathcal{L}}=100~fb^{-1}$ $yr^{-1}$}.

\subsection{Angular distribution of muon pair production $e^- e^+ \to \mu^- \mu^+$ in the NCSM }
The angular distribution of the final state scattered particles is a useful tool to understand the nature of new physics. We will now see how the azimuthal distribution of the final state scattered particles can be used to separate out the noncommutative geometry, the NCSM, from the other type of new physics models e.g supersymmetry, brane-world gravity, unparticle scenario, little Higgs models etc. 
\begin{figure}[htbp]
\vspace{-1.15in}
\centerline{\hspace{-12.3mm}
{\epsfxsize=9cm\epsfbox{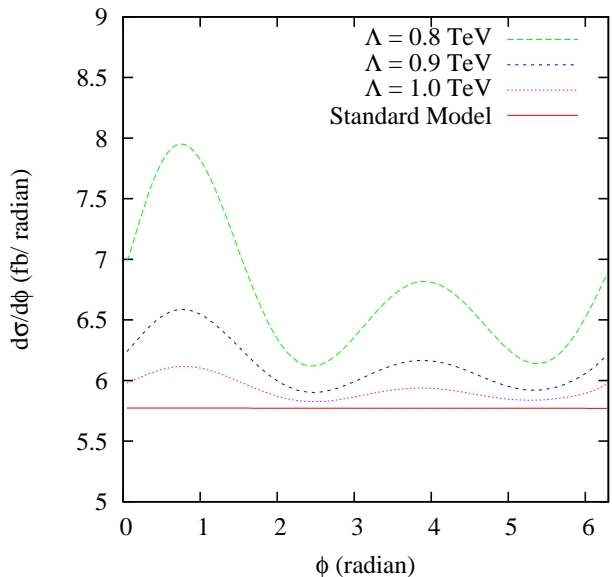}} }
\vspace{-1.25in}
\caption{{The $ \frac{d\sigma}{d\phi} $($fb/rad$) distribution as a function of $\phi$(in rad). The machine energy $E_{com}(=\sqrt{s}$) is fixed at $1.75$ TeV. The lowest horizontal curve is due to the SM, whereas the plots above the horizontal one, as we move up correspond to $\Lambda = 1.0, 0.9,$ and $0.8$ TeV, respectively.}}
\protect\label{dsdphiplot}
\end{figure}

In Fig. \ref{dsdphiplot} we show $\frac{d\sigma}{d\phi}$ as a function of the azimuthal angle $\phi$.  For the angular analysis study, we fixed the machine energy $E_{com}(=\sqrt{s}$) at $1.75$ TeV. The standard model which  is completely $\phi$ symmetric, predicts a flat distribution for $d\sigma/d\phi$. The lowest horizontal curve establishes this fact. Other plots above the horizontal one, as we move up, correspond to $\Lambda = 1.0, 0.9 $ and $0.8$ TeV, respectively in the NCSM.  The departure from the flat behavior is due to $ p_2 \Theta p_1$ and $p_4 \Theta p_3$ terms in Eqs. \ref{gamma} and \ref{Z} that bring in the $\phi$ dependence which is thus observed in Fig. \ref{dsdphiplot}. Interestingly, the curves show several maxima and minima. The largest maxima for each of the three curves is peaked at  $\phi = 0.78$ rad, whereas the second largest maxima is found to be located at $\phi \sim 4$ rad. Two minimas are found: they are located at $\phi = 2.6$ rad and $5.3$ rad, respectively. Note that in each of the above three plots, if we set $\Lambda = \infty$, the lowest horizontal SM curve is recovered. It is worthwhile to note that such an azimuthal distribution clearly reflects the exclusive nature of the noncommutative geometry which is rarely to be found in other types/classes of new physics models and hence may serve as Occam's razor- either selecting or ruling out a class of new physics model(s). 

\begin{figure}[htbp]
\vspace{-1.15in}
\centerline{\hspace{-14.3mm}
{\epsfxsize=9cm\epsfbox{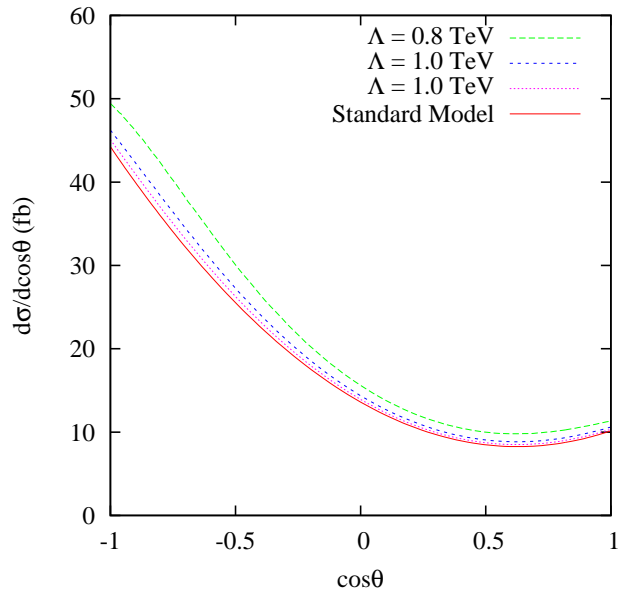}}}
\vspace{-1.15in}
\caption{The $ \frac{d\sigma}{dcos\theta} $($fb$) distribution as a function of $cos\theta$ is shown. The machine energy $E_{com}(=\sqrt{s}$) is fixed at $1.75$ TeV. The lowermost curve corresponds to the CSM, whereas the plots above the horizontal one, as we move up, correspond to $\Lambda = 1.0, 0.9,$ and $0.8$ TeV, respectively in the NCSM.}
\protect\label{dsdcosthetaplot}
\end{figure}

 We next analyze the polar distribution. In Fig. \ref{dsdcosthetaplot}, $\frac{d\sigma}{dcos\theta}$ is plotted as a function of $cos\theta$ with the machine energy $E_{com}$ being fixed at $1.75$ TeV.
Note the asymmetry of the distribution around the $cos\theta = 0$ line of Fig. \ref{dsdcosthetaplot}.  The lowermost plot in Fig. \ref{dsdcosthetaplot} corresponds to the standard model and the plots, as we move up correspond to $\Lambda = 1.0, 0.9,$ and $0.8$ TeV, respectively. 
 The uppermost curve in the figure corresponding to $\Lambda = 0.8$ TeV exhibits maximal deviation from the lowermost CSM curve (obtained by setting $\Lambda \to \infty$).

\section{Conclusion}
The idea that around the TeV scale the space and time coordinates become noncommutative(i.e. no longer commutative in nature) draws a lot of attention in the physics community. We explored the impact of space-time noncommutativity in the fundamental processes $e^+ e^- \to \gamma, Z \to  \mu^+ \mu^-$. Interestingly, we found that when we use the ${\mathcal{O}}(\Theta^2)$ Feynman rules, the 
${\mathcal{O}}(\Theta)$, ${\mathcal{O}}(\Theta^2)$ and ${\mathcal{O}}(\Theta^3)$ contributions to the cross section simply get canceled and the lowest order contribution to the cross section appears at ${\mathcal{O}}(\Theta^4)$.  We made our analyses to this order. 
The plots showing the total cross section as a function of the machine energy $E_{com}$ establish the fact that at and above $E_{com} \ge 1400$ GeV, one can expect to see the effect of noncommutative geometry at a linear collider. Setting the ILC energy at $E_{com} = 1750$ GeV, the NC scale $\Lambda = 0.8$ TeV and assuming an integrated luminosity about $100~fb^{-1}$, we estimate the signal (NC event) as about $4200$ per year and the background as (SM event) $3600$ per year. 
The azimuthal distribution $d\sigma/d\phi$, completely $\phi$ symmetric (flat) in the SM, deviates substantially in the NCSM. The deviation increases as the NC scale 
$\Lambda$ decreases. Such a nontrivial azimuthal distribution is a unique feature of the NCSM and is quite uncommon in other classes of new physics models. We also study the $d\sigma/dcos\theta$ distribution as a function of $cos\theta$. Clearly, the asymmetry around the $cos\theta = 0$ curve persists even when space-time is noncommutative.  Thus the noncommutative geometry is quite rich in terms of its phenomenological implications and it is worthwhile to explore several other interesting processes, potentially relevant for the future International Linear Collider. 
\vspace*{-0.15in}
\begin{acknowledgments}
The work of P.K.Das is supported by the DST Fast Track Project No. SR/FTP/PS-11/2006. 
\end{acknowledgments}
\vspace*{-0.15in}
\appendix

\section{Feynman rules to order ${\mathcal{O}}(\Theta^2)$ }
\noindent Following Ref. \cite{Ana}, the Feynman rule for the $f (p_{in})- f(p_{out}) + \gamma(k)$ vertex (where $f$ represents a fermion) to $O(\Theta^2)$ can be written as $ i g V_\mu(p_{out}, k , p_{in})$, where
\bea
V_\mu^{(1)}(p_{out},k,p_{in}) = \frac{i}{2} [(k\Theta)_\mu \not{p_{in}} (1 - 4 c_\psi^{(1)}) + 2 (k\Theta)_\mu (c_A^{(1)} - c_\psi^{(1)}) - (p_{in} \Theta)_\mu \not{k} \nonumber \\ - (k \Theta p_{in})\gamma_\mu) ]. \\
V_\mu^{(2)}(p_{out},k,p_{in}) = \frac{1}{8} (k \Theta p_{in}) [(k\Theta)_\mu \not{p_{in}} (1 - 16 c_\psi^{(2)}) + 4 (k\Theta)_\mu (c_A^{(1)} - 2 c_\psi^{(2)}) \nonumber \\ - (p_{in} \Theta)_\mu \not{k} - (k \Theta p_{in})\gamma_\mu) ].
\eea

\noindent The same for the vertex $f(p_{in})- f(p_{out}) + Z(k)$ can be obtained by making the necessary substitutions of $\gamma_\mu \longrightarrow g_V \gamma_\mu - g_A \gamma_\mu \gamma_5 $.
Here $\Theta_{\mu \nu \rho}= \Theta_{\mu \nu} \gamma_\rho + \Theta_{\nu \rho}
\gamma_\mu + \Theta_{\rho \mu} \gamma_\nu$, $ p_{out} \Theta p_{in} = p_{out}^\mu  \Theta_{\mu \nu}  p_{in}^\nu = -p_{in} \Theta p_{out}$. In our analysis, we set $c_A^{(1)}=c_\psi^{(1)}=c_\psi^{(2)}=0$.

\noindent The momentum conservation reads as $p_{in} + k = p_{out}$. 

\section{Momentum prescriptions and dot products}
Working in the center of momentum frame and ignoring electron and muon masses, we can specify the 4 momenta of the particles as follows:
\bea
\label{prescstart} p_1 &=& \left(\frac{\sqrt{s}}{2}, 0, 0, \frac{\sqrt{s}}{2}\right)\\
p_2 &=& \left(\frac{\sqrt{s}}{2}, 0, 0, -\frac{\sqrt{s}}{2}\right)\\
p_3 &=& \left(\frac{\sqrt{s}}{2},\frac{\sqrt{s}}{2} \sin\theta \cos\phi,\frac{\sqrt{s}}{2} \sin\theta \sin\phi, \frac{\sqrt{s}}{2} \cos\theta \right)   \\
p_4 &=& \left(\frac{\sqrt{s}}{2},-\frac{\sqrt{s}}{2} \sin\theta \cos\phi,-\frac{\sqrt{s}}{2} \sin\theta \sin\phi, -\frac{\sqrt{s}}{2} \cos\theta \right),
\eea
where $\theta$ is the scattering angle made by the $3$-momentum vector $p_3$ of $\mu^-(p_3)$ with the +ve Z axis and $\phi$ is the azimuthal angle. 
We note that the antisymmetric $\Theta_{\mu \nu}$ has $6$ independent components corresponding to $c_{\mu \nu} = (c_{0i}, c_{ij})$ with $i, j = 1,2,3$. Assuming all of them are nonvanishing they can be written in the form
\bea
c_{0i} &=& \frac{\xi_i}{\Lambda^2}, \\
\label{prescend}
c_{ij} &=& \frac{\epsilon_{ijk} \chi^k}{\Lambda^2}.
\eea
The antisymmetric $\Theta_{\mu \nu}$ is analogous to the field tensor $F_{\mu \nu}$ where $\xi_i$ and $\chi_i$ are like the components of the electric and magnetic field vectors.
Setting $\xi_i=(\vec{E})_i = \frac{1}{\sqrt{3}},~i = 1, 2, 3$ and $\chi_i= (\vec{B})_i = \frac{1}{\sqrt{3}},~i = 1, 2, 3$( noting the fact that $\chi_i = - \chi^i$, $\xi_i = - \xi^i$ and $\xi_i \xi^j = \frac{1}{3} \delta_i^j$ and $\chi_i \chi^j = \frac{1}{3} \delta_i^j$, we find
\bea
p_2 \Theta p_1 &=&  \frac{s}{2 \sqrt{3} \Lambda^2}, \\
p_4 \Theta p_3 &=&  \frac{s}{2\sqrt{3} \Lambda^2} \left[\cos\theta + \sin\theta (\cos\phi + \sin\phi) \right].
\eea
\section{Spin-averaged squared-amplitude for $ e^+ e^- \longrightarrow \mu^+ \mu^-$}
\noindent The squared-amplitude terms of Eq.\ref{Ampsq} are
\beq
\overline {|{\mathcal{A}}|^2} = {\overline {|{\mathcal{A}}_\gamma|^2}} + {\overline {|{\mathcal{A}}_Z|^2}} + 2 {\overline {Re({\mathcal{A}}_\gamma {\mathcal{A}}_{Z \dagger })}} = \frac{1}{4} |{\mathcal{A}}|^2 . 
\eeq

\noindent The various components of the squared-matrix element [Eq. \ref{Ampsq}] are found to be
\bea
{\overline {|{\mathcal{A}}_\gamma|^2}} &=& \frac{128 \pi^2 \alpha^2~A_{NC}}{s^2}  
\left[(p_1.p_3)(p_2.p_4)+(p_1.p_4)(p_2.p_3)\right],
\label{ampgamma} 
\eea
\bea
{\overline{|{\mathcal{A}}_Z|^2}} &=& \frac{8 \pi^2 \alpha^2 ~A_{NC}}{\sin^4 (2\theta_W)~s_Z^2}   
\left[T_1 (p_1.p_4)(p_2.p_3) + T_2 (p_1.p_3)(p_2.p_4) \right],  
\label{ampZ}
\eea
\bea
2 Re (\overline{A_Z A_\gamma^\dagger}) &=& \frac{64 \pi^2 \alpha^2~A_{NC}}{\sin^2 (2\theta_W)} \frac{1}{s} \frac{(s-m_Z^2)}{s_Z^2} \left[T_3 (p_1.p_3)(p_2.p_4) + T_4 (p_1.p_4)(p_2.p_3) \right], 
\label{ampinter}
\eea
where  $s_Z^2 = [(s-m_Z^2)^2 + \Gamma_Z^2 m_Z^2]$, $T_1 = 1 + 6 (4 s_W^2 - 1)^2 + (4 s_W^2 - 1)^4$, $T_2 = 1 - 2 (4 s_W^2 - 1)^2 + (4 s_W^2 - 1)^4$, ~$T_3 = -1 + (4 s_W^2 - 1)^2 $ and $T_4 = (4 s_W^2 - 1)^2  + 1$. 
\noindent In the above, one finds 
\bea
A_{NC} = \left[(1- \frac{(p_2 \Theta p_1)^2}{8})^2 + \frac{(p_2 \Theta p_1)^2}{4}\right] \left[(1- \frac{(p_4 \Theta p_3)^2}{8})^2 + \frac{(p_4 \Theta p_3)^2}{4}\right]\nonumber \\
= \left[1 + \frac{(p_2 \Theta p_1)^4}{64} \right] \left[1 + \frac{(p_4 \Theta p_3)^4}{64} \right]
\eea 
Interestingly, the inclusion of the order ${\mathcal{O}}(\Theta^2)$ Feynman rule leads to the cancellation of all lower order terms. A relevant discussion is made at the end of this appendix and in the next appendix.

\noindent Equations \ref{ampgamma} - \ref{ampinter} can be rewritten using the results and prescriptions of Appendix B as 
\bea
{\overline {|{\mathcal{A}}_\gamma|^2}} = 16 \pi^2 \alpha^2 (1 + \cos^2\theta) \times A_{NC} 
\eea
\bea
{\overline{|{\mathcal{A}}_Z|^2}} = \frac{4 \pi^2 \alpha^2}{\sin^4(2\theta_W)} \frac{s^2}{[(s-m_Z^2)+\Gamma_Z^2m_Z^2]} [(1 - 4 s^2_W + 8 s^4_W)^2 (1 + \cos^2\theta) + 2 (1 - 4 s^2_W)^2 \cos\theta] \times A_{NC} \nonumber \\ \\
2 Re (\overline{A_Z A_\gamma^\dagger}) = \frac{8 \pi^2 \alpha^2}{\sin^2 (2\theta_W)} \frac{s(s-m_Z^2)}{[(s-m_Z^2)^2 + \Gamma_Z^2 m_Z^2]} [(1 - 4 s^2_W)^2 (1 + \cos^2\theta) + 2 \cos\theta ] \times A_{NC} \nonumber \\ 
\eea
where $s_W = \sin \theta_W$.
 
\begin{flushleft}
\section {A note on the order by order (in $\Theta$) vanishing contribution to the cross section }
\end{flushleft}

The process that we are considering is a single $s$-channel process and interestingly the inclusion of 
$\mathcal{O}(\Theta^2)$ terms in the matter-gauge boson interaction vertex, leads to the cancellation of all lower order ${\mathcal{O}}(\Theta,~\Theta^2,~\Theta^3)$ terms in the cross section. The contribution (nonvanishing) to the cross section discussed so far is $\mathcal{O}(\Theta^4)$ ($ \mathcal{O}(1/\Lambda^8)$). It should be recognized that an accurate description, as a result of the cancellation would now require Feynman rules to all orders. In fact, it is quite likely that such cancellations successively continue even with the higher order terms, ultimately resulting in no net  effect from space-timenoncommutativity! This was shown in a work by Hewett \etal \cite{Hewett01} where the effect merely pulls out as an overall phase factor. However, these remain only speculative due to the unavailability and inherent difficulty in generating higher order Feynman rules. In any case, we believe that the work presented in this paper throws light on crucial aspects with regard to NCSM-- not only in demonstrating the nature of subtle modifications in the distribution curves, but also in demonstrating that such effects in a single channel process can in fact successively negate, resulting in no net effect and rendering the traditional perturbative approach ineffective.


\end{document}